\documentclass[twocolumn]{aastex631}
\usepackage{hyperref}
\shorttitle{BBH formation channels}
\shortauthors{Mukherjee and Moradinezhad Dizgah}

\begin{document}

\title{Towards a Precision Measurement of Binary Black Holes Formation Channels Using Gravitational Waves and Emission Lines}

\correspondingauthor{Suvodip Mukherjee}
\email{{smukherjee1@perimeterinstitute.ca}, Azadeh.MoradinezhadDizgah@unige.ch}

\author[0000-0002-3373-5236]{Suvodip Mukherjee}
\affiliation{Perimeter Institute for Theoretical Physics, 31 Caroline Street N., Waterloo, Ontario, N2L 2Y5, Canada}

\author[0000-0001-8841-9989]{Azadeh Moradinezhad Dizgah}
\affiliation{D\'epartement de Physique Th\'eorique, Universit\'e de Gen\`eve, 24 quai Ernest Ansermet, 1211 Gen\`eve 4, Switzerland}



\begin{abstract}

The formation of compact objects---neutron stars, black holes, and supermassive black holes---and its connection to the chemical composition of the galaxies is one of the central questions in astrophysics. We propose a novel \textit{data-driven}, multi-messenger technique to address this question by exploiting the inevitable correlation between gravitational waves and atomic/molecular emission line signals. For a fiducial probability distribution function $p(t_d)\propto t_d^{-\kappa}$ of time delays, this method can provide a measurement of the minimum delay time of $0.5$ Gyr and the power-law index $\kappa=1$ with a standard deviation $0.12$ (and $0.45$) and $0.06$ (and $0.34$), respectively from five years of LIGO-Virgo-KAGRA observation in synergy with SPHEREx line intensity mapping (and DESI emission-line galaxies). Such measurements will provide data-driven multi-messenger constraints on the delay time distribution which is currently not well known.   


\end{abstract}



\section{Introduction} Gravitational waves discovered by the LIGO-Virgo-KAGRA (LVK) scientific collaboration \citep{LIGOScientific:2014pky,Martynov:2016fzi,Acernese_2014,PhysRevLett.123.231108,KAGRA:2013pob,Akutsu:2018axf,KAGRA:2020tym} have opened a new window to explore the cosmos using coalescing compact objects such as binary neutron stars (BNS), binary black holes (BBH), and neutron star - black hole (NSBH) systems \citep{PhysRevLett.116.061102,TheLIGOScientific:2017qsa,LIGOScientific:2020kqk,Abbott_2021, LIGOScientific:2021psn,LIGOScientific:2021djp,Virgo:2021bbr}. One of the key questions that these measurements can shed light on is the population of binary compact objects and the evolution of their merger rate with cosmic time \citep{Fishbach:2018edt, LIGOScientific:2018jsj, LIGOScientific:2020kqk, LIGOScientific:2021psn}. 
The merger rate of BBH as a function of redshift, inferred until O3 LVK observation, is in agreement with a power-law form $(1+z)^\gamma$, with $\gamma=2.7_{-1.7}^{+1.8}$ for $z<1$ \citep{LIGOScientific:2021psn}. 

The observed population of GW sources can provide insight into how these binary compact objects form in the Universe and how their formation relates to the stellar population through cosmic star formation rate (SFR), stellar mass, and stellar metallicity. While the dependence of the binary merger population on these quantities is not yet known from the data, several simulations have explored the connection between stellar properties and different types of compact objects such as BNS, NSBH, and BBHs \citep{Fishbach:2018edt,Artale:2019doq, Artale:2019tfl}. However, inference of the connection of GW compact objects with host galaxies from observations is challenging due to the large sky localization error of the GW sources. 

We propose a novel multi-messenger technique to probe this connection by taking advantage of synergies between atomic and molecular emission lines such as H$\alpha$, H$\beta$, [OII], [OIII], [NII], CO, [CII] as tracers of the stellar properties and GW sources such as BHS, NSBH, and BBHs as tracers of compact objects. The spectral lines can be detected either in individually resolved galaxies by spectroscopic galaxy surveys probing emission-line galaxies (ELGs) or as aggregate signals using the emerging technique of line intensity mapping (LIM) \citep{Kovetz:2017agg}. In the latter, the spatial fluctuations of the intensity of the emission line together with its frequency provide a three-dimensional map of the cosmic large-scale structure. The dependencies of the emission lines and GW sources on stellar populations lead to an inevitable correlation between the two sectors, which depends on the formation channels of GW sources. The luminosity of emission lines determines the amplitude and redshift-dependence of the line power spectrum signal measured by LIM surveys, and the mean number densities of galaxies probed by galaxy surveys. Therefore, the correlation between the observations in GWs and emission lines can provide a model-independent measurement of the dependence of the compact object population on the SFR, stellar mass, and metallicity. Since the proposed method does not rely on cross-correlation of the GW sources with EM probes, the results are not susceptible to poor sky localization error of the GW sources. For GW sources with very good sky localization, spatial cross-correlation with EM probes to explore the delay time would provide additional constraining power \citep{ Mukherjee:2020hyn,Diaz:2021pem,Scelfo:2021fqe}.


\section{Relating GW sources and galaxy properties}\label{sec:gw} Several formation channels for binary compact systems such as BNS, BBH, and NSBH have been proposed in the literature \citep{2010ApJ...716..615O,2010MNRAS.402..371B, 2012ApJ...759...52D}. Different channels lead to different delay time distributions (DTD) $P(t_d)$, which capture the time between the formation of stars and the merging of compact objects. The DTD leaves an imprint on the merger rates of binary systems, their redshift evolution, and their connection with the host galaxies. The merger rate of the binary compact objects can be written in terms of the DTD $P(t_d)$, and the SFR density $R_{\rm SFR}(z)$ as
\begin{equation}\label{a1}
    N_{\rm GW}(z_m)= \mathcal{N}\int^{\infty}_{z_m} dz \frac{dt_f}{dz}P(t_d) R_{\rm SFR}(z),
\end{equation}
where $\mathcal{N}$ is the normalization constant to set the local merger rate of GW sources with the observed value \citep{LIGOScientific:2021psn,LIGOScientific:2021djp,Virgo:2021bbr}. The probability distribution of the delay time is expected to follow a power-law form, $P(t_d)= t_d^{-\kappa}$, with the value of $\kappa=1$ for an initial logarithmic distribution of the separation of the compact objects \citep{McCarthy:2020jwq}.  
Recent works \citep{Toffano:2019ekp, Artale:2019tfl} based on population synthesis simulations combined with galaxy catalogs from hydrodynamical simulations have shown that a different DTD of the GW sources can lead to a different dependence of the compact object mergers on the stellar mass ($M_\star$), SFR, and stellar metallicity ($Z$) of the host galaxy, which can be expressed  by the relation \citep{Artale:2019tfl}
\begin{eqnarray}\label{gwsfr}
    \log({n_{\rm GW}/{\rm Gyr}^{-1}}) &=& \gamma_1\log({M_*/M_\odot})+ \gamma_2\log({Z/Z_\odot}) \nonumber \\
    &+&\gamma_3\log({{\rm SFR}/M_\odot {\rm yr}^{-1}}) +\gamma_4.
\end{eqnarray}

\begin{figure}
    \centering
    \includegraphics[width = 0.43\textwidth]{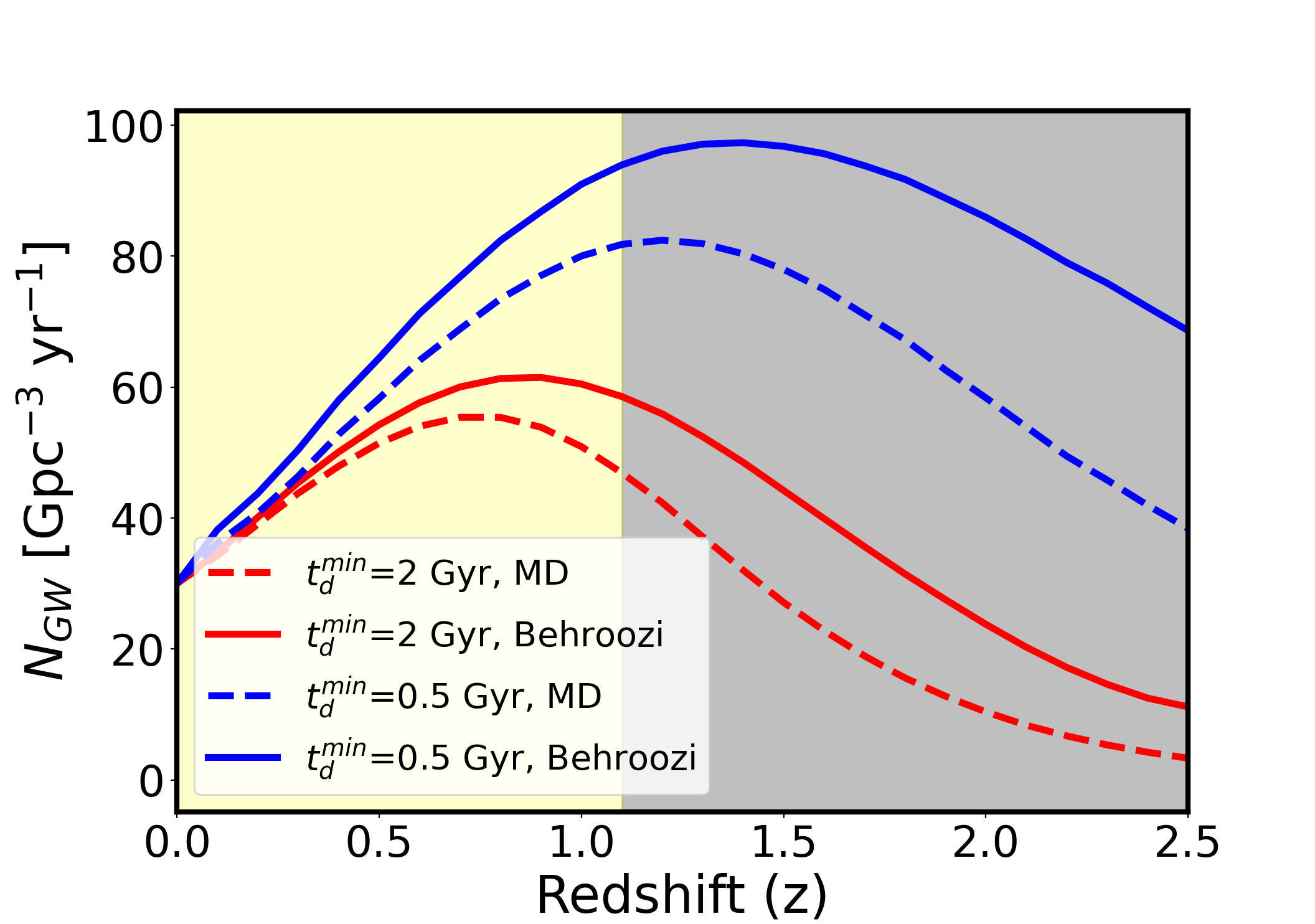}
    \caption{The merger rate of GW sources for Madau-Dickinson (MD) and Behroozi models of SFR for two different values of the minimum delay time $t_d^{\rm min}=0.5$ Gyr and 2 Gyrs. The yellow shaded region denotes the maximum redshift range up to which GW sources can be detected by the LVK detectors. The grey-shaded region is accessible by the stochastic GW background.}
    \label{fig:rates}
\end{figure}
Currently, there is no direct measurement confirming this relation from the LVK data. Recent studies from individual events \citep{Fishbach:2021mhp} and the stochastic GW background \citep{Mukherjee:2021ags} have only been able to put an upper bound on the delay time, assuming a fixed SFR. An upper bound on the stochastic GW background also puts a limit on the high redshift merger rate \citep{KAGRA:2021kbb}. The measurement of the DTD will be a direct probe to understand the dependence of the binary compact objects on stellar properties. The third-generation GW detectors such as Cosmic Explorer (CE) \citep{2019CQGra..36v5002H} and Einstein Telescope (ET) \citep{Sathyaprakash:2009xt, Punturo:2010zz}, are expected to have the sensitivity that can allow the inference of the SFR \citep{Vitale:2018yhm} from GW observations. In Fig. \ref{fig:rates}, we show the merger rate of the GW sources for two values of the minimum delay time $t_d^{\rm min}$ (in red and blue), assuming two SFR models; Madau-Dickinson \citep{Madau2014} (dashed) and Behroozi \citep{Behroozi:2012sp} (solid). A different SFR model results in a variation of the redshift evolution of the merger rate, even if assuming the same form of DTD. 

The DTD for BNS has been studied in previous works \citep{2019ApJ...878L..12S, 2019ApJ...878L..14S, 2020ApJ...905...21A}. Our focus in this work is on the distribution of delay times for BBHs. 


\section{EM probes of the galaxy properties} 
In general, the luminosities of the emission lines trace the star formation and are expected to be correlated with the astrophysical properties of the galaxies. In analogy with Eq. \ref{gwsfr}, this relation can be parameterized as 
\begin{eqnarray}\label{linessfr}
    \log{(L_{\rm line}/L_\odot)} &=& \alpha_1\log({M_*/M_\odot})+ \alpha_2\log({Z/Z_\odot}) \nonumber \\
    &+&\alpha_3\log({{\rm SFR}/M_\odot {\rm yr}^{-1}}) +\alpha_4,
\end{eqnarray}
where, $\alpha_n$ are the unknown coefficients of the fit, which vary for different lines. For instance, while [OIII], [OII], and [CII] lines are expected to have a strong dependence on metallicity, H$\alpha$, and H$\beta$ lines should not \citep{Favole:2019ouk,Yang2020}.

For the forecasts presented in this paper, in the absence of detailed models of line luminosities as in Eq. \ref{linessfr}, we use the existing scaling relations \citep{Kennicutt:1998zb, Ly:2006hx, Silva:2014ira, Gong_2017} to relate $L_{\rm line}$ and SFR, assuming them to depend only on halo mass and redshift. Therefore, ${\rm SFR}(M,z) = \Gamma_{\rm line} \times L_{{\rm line}}(M,z)$, with $\Gamma_{\rm line}$ being the scaling coefficient for a given line. We use the empirical fit from \cite{Behroozi:2012sp} for ${\rm SFR}(M,z)$. To have redshift overlap with LVK observations ($z\leq 1.1$), we consider H$\alpha$ 6563{\AA}, [OIII] 5007{\AA}, and [OII] 3727{\AA} probed by SPHEREx LIM, and the mean number densities of [OII] ELGs detected by DESI. As \cite{Gong_2017}, taking SFR in unit of $[M_\odot {\rm yr}^{-1}]$ and luminosity in unit of $[{\rm erg}\ s^{-1}]$, we consider $\Gamma_{\rm line} = (7.9, 7.6, 14) \times 10^{-42}$ for H$\alpha$, [OIII], and [OII] lines, respectively.

 {Given a model of line luminosity, we can compute the mean number density of individually resolved line emitters, $\bar n (z) = \int_{M_{\rm min}}^{M_{\rm max}} dM \ n(M,z)$, by matching the minimum mass ($M_{\rm min}$) of halos hosting line-luminous galaxies to the minimum luminosity ($L_{\rm min}$) detectable by a given galaxy survey. Considering the [OII] galaxies observed by the upcoming DESI survey}, we set $\log_{10}(L_{\rm min}/{\rm erg \ s^{-1}}) = (40.5,41)$ for redshift ranges $0.65 \leq z< 1$, and $1 \leq z \leq 1.65$, which are in broad in agreement with values in \cite{Comparat:2016jqq} and \cite{Saito:2020qxq}.

For LIM, we are interested in the statistical properties of the fluctuations of line intensities. We summarize the key quantities and refer to  \cite{Gong:2020lim} and \cite{MoradinezhadDizgah:2021upg} for further details. Using the halo-model framework, the mean intensity of a line (in the unit of $[{\rm Jys/sr}]$) is related to DM halo properties as
\begin{equation}
     {\bar I}_{\rm line}(z) = \int_{M_{\rm min}}^{M_{\rm max}} dM \  n(M,z) \frac{L_{\rm line}(M,z)}{4 \pi \mathcal D_L^2(z)} \left (\frac{dl}{d\theta} \right )^{2} \frac{dl}{d\nu},
\end{equation}
where $n(M,z)$ is the halo mass function, $\mathcal D_L$ is the luminosity distance, and $(dl/d\theta)^2 dl/d\nu$ reflects the conversion from comoving volume to observed specific intensity volume defined in terms frequency, $\nu$, and angular size, $\theta$. 
On large scales, line intensity fluctuations can be described as a linearly biased tracer of the underlying DM distribution, $\delta_{\rm line}({\bf k},z) = \bar I_{\rm line}(z) b_{\rm line}(z) \delta_m({\bf k},z)$. Therefore, in the absence of anisotropies, the clustering component of the line intensity power spectrum (in the unit of [$({\rm Jy/sr})^2 ({\rm Mpc/h})^3$]) is given by
\begin{equation}\label{eq:pk_clust} 
P_{\rm LIM}^{\rm clust}(k,z) =  \left[{\bar I}_{\rm line}(z) b_{\rm line}(z)\right]^2 P_m(k, z),
\end{equation}
Additionally, the observed line power spectrum includes a shot-noise contribution due to the discreteness of line-emitting galaxies and the thermal noise of the instrument. In the Poisson limit, the former is given by
\begin{equation}\label{eq:pk_shot}
\hspace{-.1in}P_{\rm LIM}^{\rm shot}(z) = \int_{M_{\rm min}}^{M_{\rm max}} dM \ n(M,z) {\left[\frac{L_{\rm line}(M,z)}{4 \pi \mathcal D_L^2(z)} \left (\frac{dl}{d\theta} \right )^{2} \frac{dl}{d\nu} \right ]}^2.
\end{equation}
The model of the power spectrum used in our forecasts further includes redshift-space distortions, Alcock-Paczynski effect, and interloper lines \citep{Lidz:2016lub, Cheng:2016yvu}. 

\section{Expected correlation between the GW signal and the emission line signal}\label{sec:gw-em} 
{As discussed in Sec. \ref{sec:gw} (Eq. \ref{gwsfr}), using numerical simulation, previous studies \citep{Artale:2019doq, Artale:2019tfl} have shown that the expected merger rates of GW sources strongly depend on the SFR, stellar mass, and metallicity. As a result, depending on the local behavior of these quantities, we can write Eq. \ref{gwsfr} in terms of a matrix} \footnote{Bold fonts denote matrices.} 
\begin{equation}\label{gwsfrmat}
   \mathbf{N} (z)= \mathbf{G}(z)\mathbf{S}(z),
\end{equation}
where $\mathbf{N} (z)= [n_{\rm BNS}(z), n_{\rm NSBH}(z), n_{\rm BBH}(z)]$ refers to the number density of GW sources of the three binary types, and $\mathbf{G}$ is a matrix of the coefficients of the relation between the number densities of GW sources and the astrophysical properties of the galaxies, represented by $\mathbf{S} (z)$. {The delay-time distribution of the compact objects controls the values of the matrix element $\mathbf{G}$ and it depends on the stellar properties of the host galaxy. 
Previous analyses by Artale {\it et al.} have shown the behavior of this scaling relation across a wide redshift range of z $\in [0,6]$. 
In Eq. \ref{gwsfrmat}, the connection of the GW merger rate to the population-III stars is not included and modification to this expression will be required.}

{In constructing Eq. \ref{gwsfrmat}, two major assumptions enter. Firstly, it depends on the theoretical modeling of the stellar properties and formation channel scenario, and this relation is not yet constrained by data. Secondly, it is written in terms of average galaxy properties at every redshift. As different galaxies at each redshift will exhibit a variation from the global average, this relation will show variations. Due to a large sky localization error of the GW sources, we cannot uniquely identify the host galaxy unless there is an EM counterpart. In this work, we propose a way to evade the first problem and partly solve the second problem by using a multi-messenger approach. In the following discussion we show that by exploring a direct relation of the GW merger rates with the signal from ELGs and LIM, we can probe their dependencies on the formation channel. Furthermore, the proposed technique makes it possible to consider halo mass-dependent SFR and also explore the dependence on metallicity and stellar mass. Therefore, we can investigate the dependence of the merger rates on different galaxy properties. In this analysis, we focus on GW sources without EM counterparts. For sources with an EM counterpart, this approach can be even more powerful.} 

{A promising observational probe of the stellar properties of the galaxies are the resolved emission line galaxies and the aggregate intensity mapping signal of spectral emission lines. The amplitudes of both signals are determined by the specific luminosity of the spectral line of interest, given in Eq. \ref{linessfr}. This relation can be cast in matrix form as}
\begin{equation}\label{linessfrmat}
\mathbf{E}(z)= \mathbf{A}(z)\mathbf{S}(z),
\end{equation}
where $\mathbf{E}(z)$ can be the intensity of the ELGs or RMS fluctuation of the LIM signal for different lines, and $\mathbf{A}$ is a matrix whose coefficients relate to the observed strength of the emission line signal with the global astrophysical properties, represented by $\mathbf{S}$. Since both GW sources and their merger rate and the line emission depend physically on the same astrophysical quantities, we can combine Eqs. \ref{gwsfrmat} and \ref{linessfrmat} as
\begin{equation}\label{linesgwmat}
\mathbf{N}(z)= \mathbf{G}(z)\mathbf{A^{-1}}(z)\mathbf{E}(z) \equiv \mathbf{K}(z)\mathbf{E}(z),
\end{equation}
where, $\mathbf{K}(z) \equiv \mathbf{G}(z)\mathbf{A^{-1}}(z)$ denotes the matrix whose coefficients capture the connection between the GW source properties and the emission line signal. In the above equation, $\mathbf{N} (z)$ and $\mathbf{E}(z)$ can be measured from GW and EM observations, respectively. 
Although the connection matrices $\mathbf{G}$ and $\mathbf{A}$ are not well known, we can infer the matrix $\mathbf{K}$ by combining the two independent observations, without invoking any model. {The values of elements of matrix $\mathbf{K}$ depend on the formation channel, which is now encapsulated in terms of the observable probes such as the ELGs and LIM signal as a function of redshift. The redshift evolution of the elements in the matrix $\mathbf{K}$ is controlled by the DTD. As we will show below, for scenarios with a large delay time, the merger rate of compact objects will show a strong correlation with a low star formation rate at a lower redshift. In comparison, if the delay time is small, then the higher merger rate of the compact objects will exhibit stronger correlations with galaxies having large star formation activity. By measuring the value of $\mathbf{K}$ as a function of redshift from a combination of GWs and emission lines data, we can reconstruct the DTD.}  {The tomographic measurement of the correlation with redshift makes it possible to reconstruct the delay time distribution.}

\begin{figure}
\centering
\includegraphics[width= 0.45\textwidth]{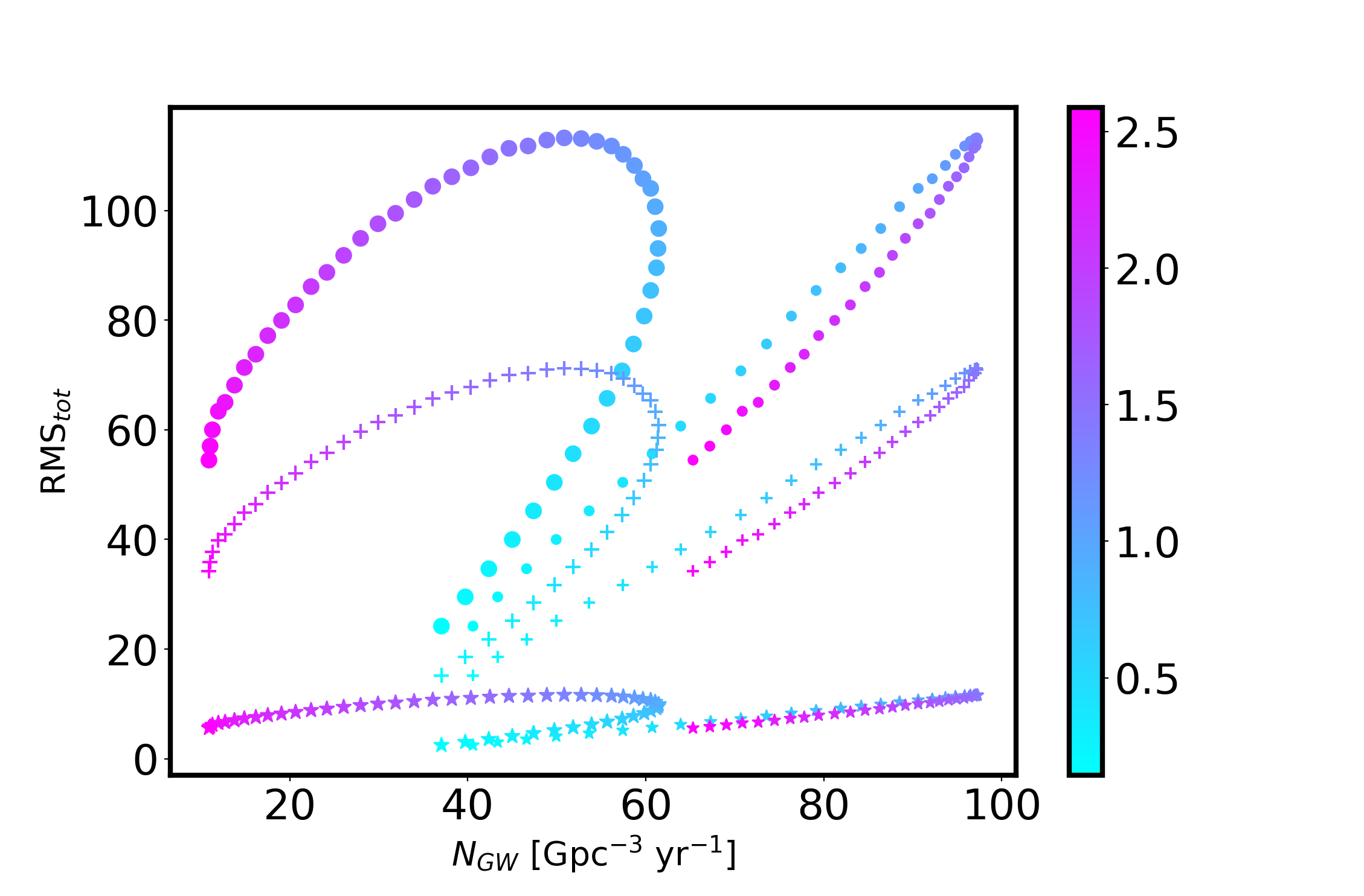}
\includegraphics[width=0.42\textwidth]{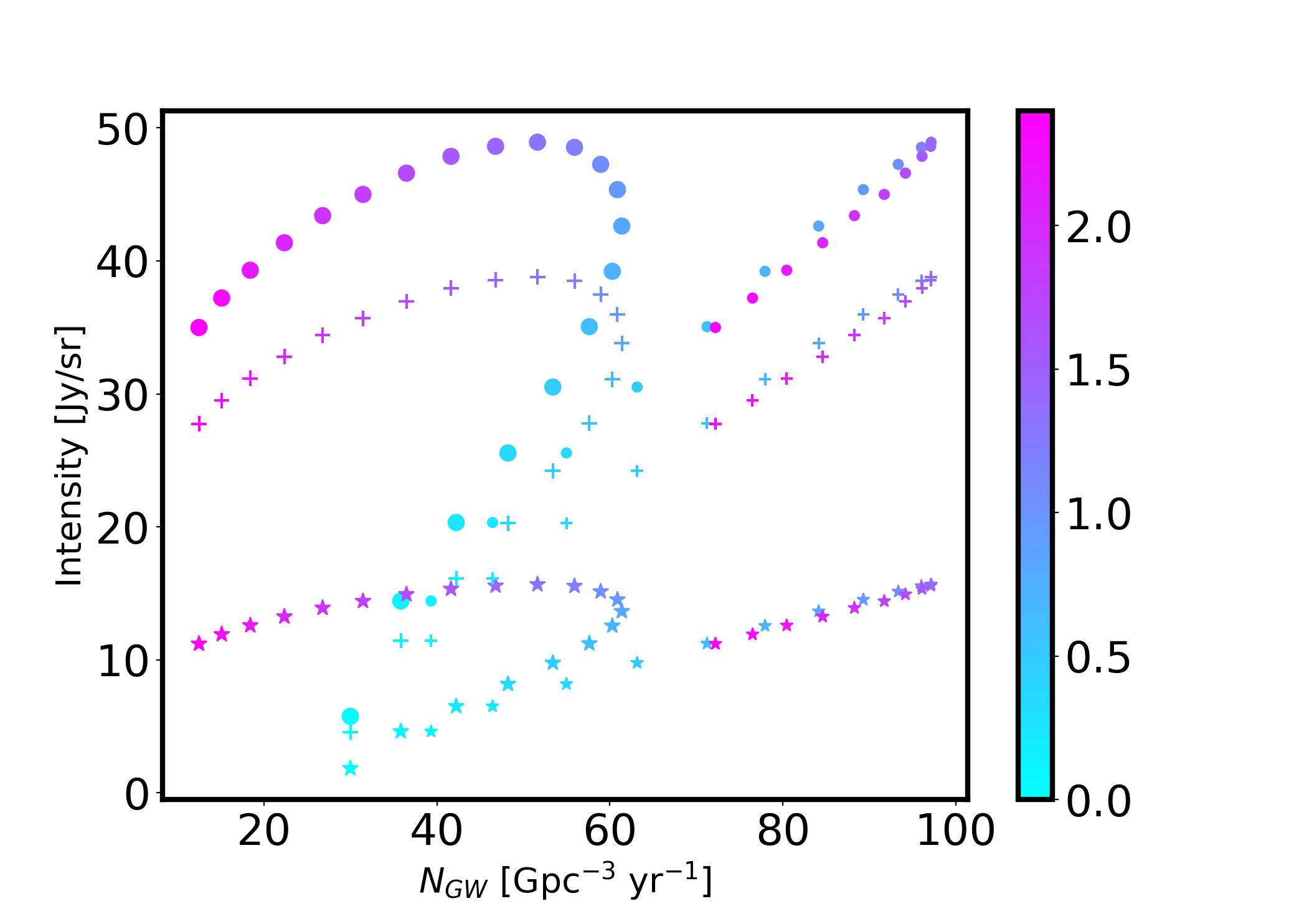}
\caption{The correlations between GW sources, N$_{\rm GW}$, and the RMS signal of LIM (top panel) and the ELG intensity (bottom panel) as a function of cosmological redshift (shown by the color bar). Different marker styles correspond to H$\alpha$ (in circle), [OIII] (in cross), and [OII] (in star) lines. The large and small markers show the correlations for two different delay time distributions; $t_d^{\rm min}= 0.5$ Gyr (small marker) and $t_d^{\rm min}=2$ Gyr (large marker).}
\vspace{.2in}
\label{fig:el-gw}
\end{figure}

{Using a generative model of GW merger rates for two scenarios of DTD and an SFR model that captures the halo mass dependence, in Fig. \ref{fig:el-gw}, we show how the correlation between the GW merger rate and ELGs (and LIM) signal of different spectral lines can vary when changing DTD. More specifically, we assume} the SFR of Behroozi {\it et al.}, and a delay time distribution of the form $P(t_d) \propto t_d^{-\kappa}$ with two different values of the minimum delay time parameter of $t^{\rm{min}}_d= 0.5$ Gyr and $2$ Gyr (shown in small- and large-size markers, respectively). The upper panel shows the correlation between N$_{\rm GW}$ and the RMS fluctuation of the LIM signal, while the lower panel shows the correlation with the mean intensity of ELGs. The three curves from top to bottom in each panel, shown by circles, crosses, and stars, correspond to H$\alpha$, [OIII], and [OII] lines. This plot indicates that the trajectory of the correlation between emission lines and GW shows different behavior for different delay time scenarios. If the delay time is small, then the GW sources will exhibit a longer correlation with the emission line signals before turnover, in comparison to the case of a large delay time. Moreover, depending on whether the strength of the emission lines signals strongly correlates at all redshifts, the correlation will show a positive helicity, in comparison to a negative helicity when the GW sources will not follow the emission lines signal up to a high redshift. It is noteworthy that GW sources of different masses can exhibit slightly different dependence on the delay time, hence a different dependence on the emission line. The probability distribution $P(t_d, \mathcal{M}) = t_d^{-\kappa(\mathcal{M})}$ with a value of minimum delay time can have a mass dependence $t_d^{\rm min}(\mathcal{M})$. Therefore, the mass dependence of the delay time can also be captured using the method presented. 

It should be noted that the redshift uncertainties in the measurements of $\mathbf{N}(z)$ and $\mathbf{E}(z)$ set the accuracy with which the redshift-dependence of $\mathbf{K}(z)$, which carries information about the formation channels of the binary compact objects, can be determined. The redshift uncertainty of the EM measurement and the luminosity distance uncertainty in the GW signal can be incorporated in the above equation with a window function, $\mathbf{W}(z,z')$ and 
$\mathbf{D}(\mathcal D_L, z)$, which modifies Eq. \ref{linesgwmat} to
\begin{eqnarray}\label{linesgwmat2}
\mathbf{N}(z)\equiv \mathbf{D}(\mathcal D_L, z)\mathbf{N}(\mathcal D_L)= \mathbf{K}(z)\mathbf{W}(z,z')\mathbf{E}(z').
\end{eqnarray}

\section{Fisher Forecasts} 
 {As described in the previous section, the merger rate of the GW sources and the number density of ELGs or the amplitude of LIM power spectrum can exhibit a correlation due to the dependence of both signals on the underlying stellar properties of the galaxies such as stellar mass, SFR, stellar metallicity. By comparing the observed redshift distribution of the ELG (or LIM) signal with the observed merger rate of the GW sources, this correlation can be measured. In this section, we apply this idea to the mock BBH samples and ELG/LIM signal generated assuming an underlying model of the SFR and delay time distribution for three different emission lines namely H$_\alpha$, O[II], and O[III]. So in Eq. \ref{linesgwmat}, the matrix $\mathbf{K}$ becomes a row matrix with dimension $1\times3$ and the matrix $\mathbf{E}$ becomes a column matrix with dimension $3\times 1$ which needs to be computed as a function of redshifts for making a tomographic estimation. By performing a Fisher matrix analysis, we show the achievable constraints on DTD} from current surveys such as LIGO \citep{LIGOScientific:2014pky,Martynov:2016fzi}, Virgo \citep{Acernese_2014,PhysRevLett.123.231108}, and KAGRA \citep{KAGRA:2013pob,Akutsu:2018axf,KAGRA:2020tym} in synergy with the near-term LIM surveys such as SPHEREx and ELGs from DESI. 


 {As described earlier, we consider \citet{Behroozi:2012sp} SFR to generate the ELGs signal as a function of redshift which is detectable by DESI using the equation}
\begin{equation}\label{elgs_nz}
\bar n (z) = \int_{M_{\rm min}}^{M_{\rm max}} dM \ n(M,z), 
\end{equation}
 {with $M_{\rm min}$ corresponding to the halo mass for the minimum luminosity detectable by surveys like DESI. For LIM signal, assuming the same SFR model, we compute the power spectrum of line intensity fluctuations (given by Eqs. \ref{eq:pk_clust} and \ref{eq:pk_shot}), and calculate the total RMS fluctuation given by}
 \begin{equation}\label{sigma_LIM}
\sigma^2_\ell(z) = \sum^{k_{\rm max}}_i k_i^2 \Delta k/(2\pi)^2 W^2(R_s,k_i) P_\ell(k_i,z).
\end{equation}
 {Here, $P_\ell(k,z) \equiv (2\ell+1)/2 \int_{-1}^1 d\mu\ P(k,\mu;z) \mathcal L_l(\mu)$, where $\mu$ is the angle with respect to line-of-sight direction, $\mathcal L_\ell$ are the Legendre polynomials (with $\ell = 0,2,4$), and $P(k,\mu;z)$ is the anisotropic power spectrum. In the summation, $k_i$ is the wavenumber at the center of the $i^{\it th}$ k-bin of width $\Delta k$ (set to be the fundamental mode of the survey), $k_{\rm max}=0.3 \ h/{\rm Mpc}$, and $W(R_s,k)$ is a top-hat smoothing kernel in real-space with $R_s=8\ {\rm Mpc}/h$.} For LIM surveys, the redshift uncertainty, $\sigma_z = (1+z)/{\mathcal R}$, is determined by the resolution of the instrument denoted by $\mathcal R \equiv \nu/{\Delta \nu}$, while for DESI, we assume $\sigma_z= 0.001(1+z)$.

 {For generating the GW sources, we consider the same \citet{Behroozi:2012sp} SFR and consider a power-law model of delay time distribution, $P(t_d) \propto t_d^{-\kappa}$, assuming three different scenarios with (i) $\kappa=1$ and a minimum delay time distribution of $t_d^{\rm min}= 0.5$ Gyr, (ii) $\kappa=1$ and $t_d^{\rm min}= 2.0$ Gyr, and (iii) $\kappa=1.2$ and $t_d^{\rm min}= 0.5$ Gyr. For each of these scenarios, we generate a mock sample of GW sources with redshift distribution using Eq. \ref{a1}.} We consider a power-law mass distribution of BBHs with ($m^{-2.35}$) for the heavier object and $m^{-1}$ for the lighter one, motivated by the recent measurements \citet{LIGOScientific:2020kqk, LIGOScientific:2021psn}. We estimate the matched filtering signal to noise ratio (SNR), $\rho$, for binary systems with detector-frame chirp mass $\mathcal{M}_d$ at the luminosity distance of $\mathcal{D}_L$, using the relation \citep{Farr:2015lna, 2021PhRvD.104f2009M}
\begin{equation}\label{snr}
    \rho= \rho_*\Theta\bigg(\frac{\mathcal{M}}{\mathcal{M}^*_d}\bigg)^{5/6}\bigg(\frac{\mathcal{D}_L^*}{\mathcal{D}_L}\bigg),
\end{equation}
where $\Theta$ is the detector projection factor assumed to be uniform between $[0,1]$.  {The parameters $\mathcal{D}_L^*$ and $\mathcal{M}^*_d$ are the reference chirp mass and luminosity distance at which an optimally oriented binary has a signal-to-noise ratio of $\rho_*=8$. {For our mock data, we set the values of $\mathcal{M}^*_d= 26.12$ M$_\odot$ and $\mathcal{D}_L^*= 2.5$ Gpc, $\mathcal{D}_L^*= 2.1$ Gpc, and $\mathcal{D}_L^*= 1.2$ Gpc for aLIGO, Virgo, and KAGRA respectively according to \cite{KAGRA:2013pob}.} 
The combined matched filtering  SNR for the network of detectors is obtained by using $\rho^2_{\rm det}= \sum_i \rho_i^2$, and we select samples with $\rho_{\rm det} \geq 10$. We draw posterior samples on the GW source parameters by using the approximate form of the likelihood for the chirp mass and mass-ratio given by \cite{Farr:2015lna} and \cite{2021PhRvD.104f2009M}. We estimate the posterior distribution of the luminosity distance using a Gaussian approximation. Though the results from individual events are likely to be non-Gaussian, the results after combining several events are expected not to vary significantly due to the approximation of the Gaussian posterior by the central limit theorem. Therefore, the Fisher forecast presented in this paper will not be significantly affected by this assumption. The detection of the GW sources of lighter mass sources situated at higher luminosity distances with $\rho_{\rm det} \geq 10$ is rare. As a result, an appropriate selection function depending on the source properties also needs to be taken into account. 

 {For the mock samples of GW sources and ELGs/LIM signals from the same generative model of SFR, using Eq. \ref{linesgwmat}, we can now infer the values of elements of matrix $\mathbf{K}$ for different emission lines as a function of redshift. {In this process, for each independent emission line, we find the values of the coefficients $K_{[GW][ELG/LIM]}$ at every redshift using the computed $N_{GW}(z)$ and ELG/LIM signals. We compute the coefficients for all the three lines considered here and obtain a matrix $\mathbf{K}$ at each redshift with an overall normalization factor that is matched with the total number of observed GW sources. The total number of GW sources cannot be constrained using this formalism (as it is absorbed in the overall normalization), but we can constrain the redshift evolution of the coefficients of the matrix element, which is useful in measuring the delay time distribution.} The values of individual coefficients are shown in Fig. \ref{fig:Kmat}, where we can see that for different delay time distributions, the elements of the matrix $\mathbf{K}$ evolve differently with redshift. As we consider only BBHs in this analysis, the corresponding matrix element for BBHs is shown as a function of redshift for three different emission lines  H$_\alpha$, O[II], and O[III].  
The overall amplitude of the distribution depends on the total number of GW sources, and its evolution with redshift is driven by the nature of delay time distribution. The correlation coefficients are larger for O[II] and O[III] lines compared to H$_\alpha$ line. {The nature of the redshift evolution of the values of the coefficients in matrix $\mathbf{K}$ is related to the correlation plot shown in Fig. \ref{fig:el-gw}. The corresponding physical explanation of this redshift evolution is discussed in Sec. \ref{sec:gw-em}. For small (or large) values of the minimum delay time and steeper (or shallower) power-law index of the delay time distribution, the mergers of the BBHs shift towards a higher (or lower) redshift. As a result, the values in the matrix $\mathbf{K}$ will show a stronger correlation with the ELGs/LIM signal for the former case than in the latter case. 
The similarity in the behavior of the values of the elements in the matrix $\mathbf{K}$ for different lines is because the same SFR model regulates the ELGs/LIM signal and for the same SFR, GW source distribution changes for different delay time distribution as a function of redshift. As the same underlying SFR  contributes to both the ELGs/LIM signal for different emission lines and GW sources, it becomes possible to infer the delay time distribution from the GW sources by jointly using both GW data and ELG/LIM data which helps in mitigating the uncertainty associated with SFR.}
For models of delay time distribution different from a power-law model, the values of the elements in the matrix $\mathbf{K}$ will show different behavior. Let us note that the delay time distribution can also depend on the source properties of the GW events such as their masses, spin, etc. We leave the investigation of how the proposed technique can be used to explore this aspect in future work.}

\begin{figure}
\centering
    \includegraphics[trim={0.cm, 0cm, 0cm, 0cm},clip,width=0.5\textwidth]{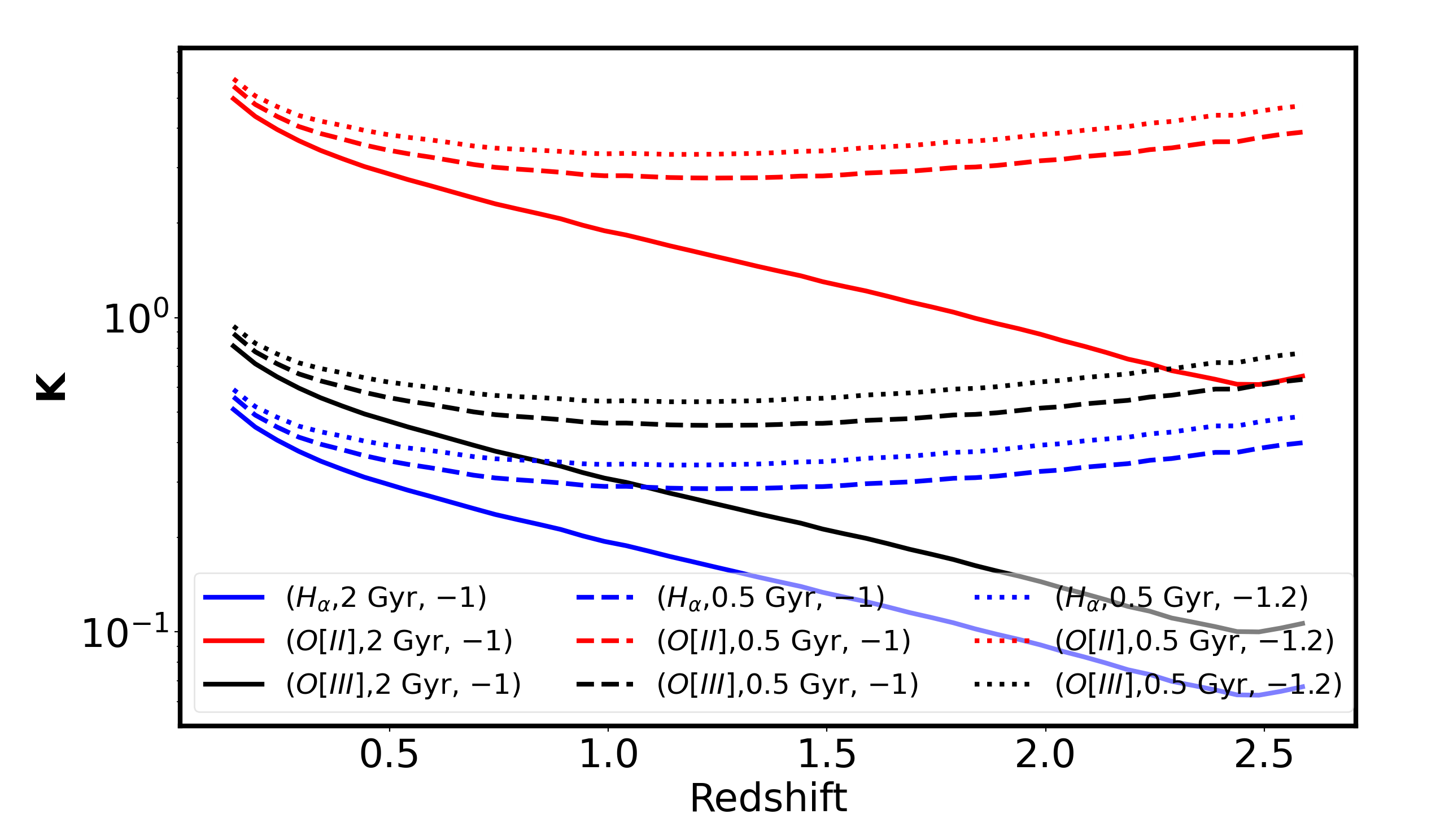}
    \caption{ {The values of elements of the matrix $\mathbf{K}$ for BBHs as a function of the redshift for different lines for three different cases of the delay time distribution indicated in the label in the following order (emission line, minimum delay time, power-law index).}} 
    \label{fig:Kmat}
\end{figure}


\begin{figure}
\centering
    \includegraphics[trim={0.3cm 0.2cm 0.2cm 0.cm},clip,width=0.4\textwidth]{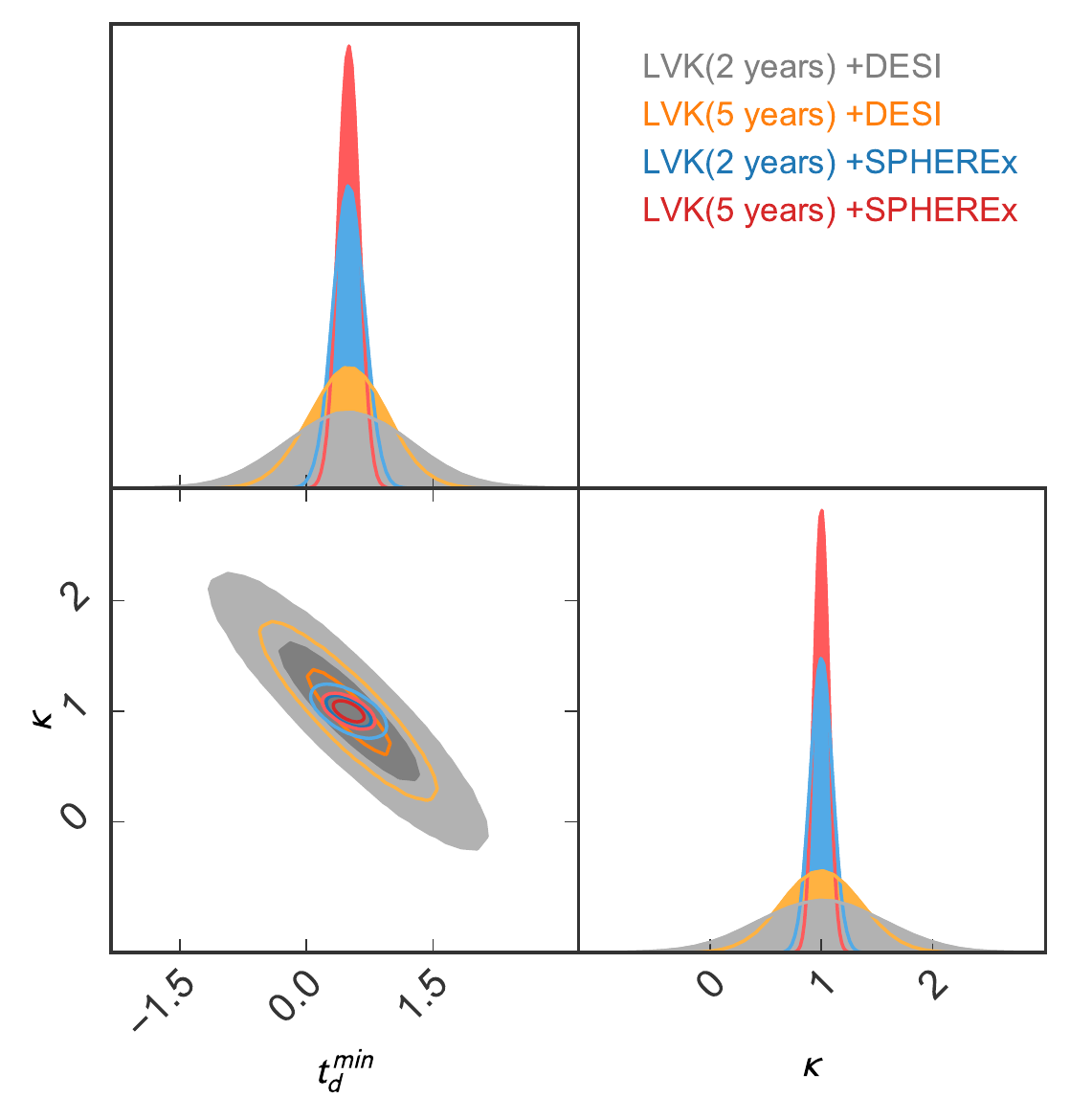}
    \caption{Fisher forecast for measuring the minimum delay time $t_d^{\rm min}=0.5$ Gyr, and power-law index $\kappa$ of the probability distribution $P(t_d)=t_d^{-\kappa}$ using LVK and the emission lines signal from LIM by combining H$\alpha$, [OIII], and [OII] lines from SPHEREx (blue and red), and [OII] ELGs from DESI (grey and yellow). The current constraints on the delay time $t_d^{\rm min}$ is $<2.2$ Gyrs at $90\%$ credible interval assuming a fixed SFR and $\kappa=1$ obtained from the individual events detected in the first half of the third observation.} 
    \label{fig:fisher}
\end{figure}
 {With the above ingredients, we perform a Fisher analysis to estimate how well the delay time distribution parameters can be measured using the correlations of GW and EM signals.} The Fisher matrix ($F_{\alpha \beta} \equiv \langle \frac{\partial^2 \mathcal{L}}{\partial\theta_\alpha \theta_\beta}\rangle$) can be written in terms of the log-likelihood ($\mathcal{L}= -\ln(L)$) for the parameters $\theta \in \{t_d^{\rm min}, \kappa\}$,
\begin{eqnarray}\label{fisher1}
2\mathcal{L}(\theta)= \sum_{L_i}\sum_{z_k} \sum_{\mathcal{M}_*}&\frac{(\hat N_{GW}(z_k,{\mathcal M}_*)- K(z_k,{\mathcal M}_*,\theta)E(z, L_i))^2}{C(z_k, {\mathcal M}_*, L_i)}\nonumber \\ & + \ln {\rm det}(C(z_k, {\mathcal M}_*, L_i)),
\end{eqnarray}
and the standard deviation (Cramer-Rao bound) on a parameter $\theta_\alpha$ denoted by $\Sigma_{\alpha}$ can be obtained by taking the square-root of the $\alpha$ component of the inverse of the Fisher matrix $\Sigma_{\alpha}= \sqrt{(\mathbf{F}^{-1})_{\alpha\alpha}}$. In Eq. \ref{fisher1}, $\hat N_{GW}(z_k,{\mathcal M}_*)$ is the observed number of GW sources at redshift $z_k$ and mass bin centered at $\mathcal{M}_*$, while $K(z_k,{\mathcal M}_*)E(z, L_i)$ denotes the theoretical prediction for the number of GW sources, $N_{GW}$, given emission line signal and different parameters related to the delay time. In the denominator, $C(z_k, {\mathcal M}_*, L_i)= \sigma^2_{p}+ \sigma^2_{e}+ \sigma^2_r$ denote the covariance matrix, which we assume to be diagonal and consisting three parts, namely $\sigma^2_{p}= N_{\rm GW}$ which is the Poisson error on the GW source number, $\sigma^2_{e}= (\partial N_{\rm GW}/\partial L_i)^2 \sigma^2_{L_i}$ is the error in the GW numbers due to change in the luminosity of the $i^{th}$ emission line, and $\sigma^2_r= (\partial N_{\rm GW}/ \partial z)^2\left[(\partial z/\partial \mathcal{D}_L)^2|_{\Theta_c} \sigma^2_{\mathcal{D}_L} + \sigma^2_z\right]$ is the error in the redshift due to the error in the luminosity distance measurement $\sigma^2_{\mathcal{D}_L}$ for a fixed cosmological parameters denoted by $\Theta_c$ (set to Planck-2018 \citep{Aghanim:2018eyx} values), and $\sigma^2_z$ is the redshift uncertainty in the EM sector.  
The major sources of uncertainty are the luminosity distance and the Poisson error on the GW sources. The additional uncertainty due to the stochasticity of the DTD specific to the sources is sub-dominant due to the all-sky averaging of the LIM signal and a large number of ELGs.  {The elements of the Fisher matrix are calculated by taking numerical derivative of the matrix elements $\mathbf{K}$ with change in the delay time distribution parameters $\theta$ as $\partial K(z_k, M_*, \theta)/\partial\theta_\alpha= \frac{1}{\Delta \theta_\alpha}[K(z_k, M_*, \theta_\alpha + \Delta \theta_\alpha/2) - K(z_k, M_*, \theta_\alpha - \Delta \theta_\alpha/2)]$.} 

For LIM,  we consider the redshift range of $z_{\rm max} = 1.1$ and $z_{\rm min} = \{0.2 , 0.5, 1.0\}$ for H$\alpha$, [OIII], and [OII] lines, which can be probed in the lowest frequency band of SPHEREx with the spectral resolution of ${\mathcal R} = 41$, and over two 100 ${\rm deg}^2$ fields \citep{Dore:2014cca}. The instrument noise $P_N(z) = V_{\rm vox} \sigma_{I, \rm vox}^2$ depends on the voxel comoving volume $V_{\rm vox}$ and the voxel intensity noise $\sigma_{I, \rm vox}$. Furthermore, we account for the attenuation of signal due to Fourier spectral and angular point-spread functions, which render the instrument noise scale-dependent.  We account for the contamination from both foreground and background line interlopers. For DESI, we consider the mean number density of the [OII] emitters at $0.65 \leq z \leq 1.1$.   

In Fig. \ref{fig:fisher}, we show the 2D marginalized constraints on the parameters $t_d^{\rm{min}}=0.5$ Gyr and $\kappa=1$ from a combination of GWs and LIM from SPHEREx and mean number densities of ELGs from DESI using the Fisher formalism. The plots indicate that we can measure delay time parameter and the fiducial model of the power-law index of the probability distribution function with corresponding error-bars of $\Sigma_{t_d^{\rm{min}}}= 0.12$ (and $0.18$) and $\Sigma_{\kappa}\sim 0.06$ (and $0.095$) for 5 years (and 2 years) of observation from LVK in synergy with SPHEREx. Using DESI, the precision on the parameters degrades by about a factor of 3.5 due to the limited availability of ELGs at lower redshift, and also for having only [OII] emission line signal. This shows that the first precise measurement of the DTD will be possible in the near future from LVK. For comparison, an upper bound on the DTD available from the first half of the third observation run is $t_d^{min}<2.2$ Gyrs \citep{Fishbach:2021mhp}, assuming a fixed value of the power-law index $\kappa=1$ and for a fixed SFR \citep{Madau:2016jbv}. The technique we propose does not require an assumption on the form of the SFR while allowing us to explore DTD and its correlation with the stellar metallicity using a multi-messenger avenue.  {Considering different SFR model parameters (with a general form given by the Madau-Dickinson model) in generating the mock data, we have found that the constraining power of the delay time distribution by this technique is possible for a wide range of SFR parameters. The model parameters which control the peak of the SFR history and the slope at the low redshift have maximum impact on the GW merger rates accessible from LVK.}

\section{Conclusions and future prospects} 

In this {\it letter}, we introduced a novel multi-messenger avenue to explore the formation channel of compact objects by studying the correlation between different emission line signals from ELGs or LIM.  
We showed that the network of LVK detectors, in conjunction with the LIM signal of H$\alpha$, [OIII] lines measured by SPHEREx or the number count of ELGs detectable by DESI, can constrain the minimum delay time value and the power-law index of the probability distribution with decent precision. The proposed method can equally be applied to understand the dependence of the compact objects on SFR and metallicity, not only for the stellar-mass black holes, but also for the BNS, NSBH, and also for SMBHs. The connection between BNS, NSBHs, and SMBHs will be possible to explore from the next generation GW detectors such as CE,  ET, and LISA in synergy with the multi-line LIM signal and ELGs.

This is the first work showing the potential of this multi-messenger approach to probe the delay time distribution of BBHs. There are several directions in which this work can be extended. This technique can further shed light on the formation channel of BBHs and can also distinguish primordial black holes (PBHs) populations. Moreover, in this current work, we have assumed a fixed value of the cosmological parameters to relate the observed GW sources with emission lines. Joint estimation of the cosmological parameters and the delay time parameter is also possible by this technique. In summary, the combined study of the emission lines and GW sources can play a crucial role in exploring several new territories in astrophysics and fundamental physics. 


{\it Acknowledgements:} The authors are thankful to Maya Fishbach for reviewing this manuscript as a part of LVK publication and presentation policy and for providing useful comments. The authors would like to thank Olivier Dor\'e, Garrett K. Keating, and Pascal Oesch for their valuable input. S.M. is supported by the Simons Foundation. Research at Perimeter Institute is supported in part by the Government of Canada through the Department of Innovation, Science and Economic Development and by the Province of Ontario through the Ministry of Colleges and Universities. A.M.D. is supported by the SNSF project ``The  Non-Gaussian  Universe and  Cosmological Symmetries", project number:200020-178787. A.M.D also acknowledges partial support from Tomalla Foundation for Gravity.  . The authors would like to thank the  LIGO/Virgo scientific collaboration for providing the noise curves. LIGO is funded by the U.S. National Science Foundation. Virgo is funded by the French Centre National de Recherche Scientifique (CNRS), the Italian Istituto Nazionale della Fisica Nucleare (INFN), and the Dutch Nikhef, with contributions by Polish and Hungarian institutes. This material is based upon work supported by NSF's LIGO Laboratory which is a major facility fully funded by the National Science Foundation.

\software{Astropy\citep{2013A&A...558A..33A}, matplotlib \citep{Hunter:2007}, NumPy \citep{2011CSE....13b..22V}, SciPy \citep{scipy}, pygtc \citep{Bocquet2016}}. 


\bibliography{main_APJL}{}
\bibliographystyle{aasjournal}



\end{document}